\documentclass[11pt,a4paper]{article}
\usepackage{makeidx}
\usepackage{epsfig}
\usepackage{amssymb}
\usepackage{mathrsfs}
\usepackage{xspace}
\usepackage[latin1]{inputenc}
\usepackage[nonstop]{feynmp}
\usepackage[english]{babel}
\def\bbra#1{\ensuremath{\Big\langle#1\Big|}}
\def\bket#1{\ensuremath{\Big|#1\Big\rangle}}
\def\D{\mathscr{D}}
\def\Phipar{\ensuremath{\Phi_\|}}
\def\Phiperp{\ensuremath{\Phi_\perp}}
\def\bra#1{\ensuremath{\left\langle#1\right|}}
\def\ket#1{\ensuremath{\left|#1\right\rangle}}
\def\d{\textrm{d}}
\def\cf{\ensuremath{C_{\!F}}}
\def\gud#1#2{\ensuremath{g^#1{}\!_#2}}

\def\phsiint#1{\ensuremath{\int\frac{\d^4#1}{(2\pi )^4i}}}
\def\ds#1{#1\!\!\!/}
\def\li{\ensuremath{\mathrm{Li}_2}}
\date{19th January 2000}
\title{Renormalon Model of Twist-4 Corrections to the Pion Distribution Amplitude}
\author{\textsc{Jeppe R.~Andersen\footnote{andersen@nbi.dk}}\\The Niels Bohr Institute\\ Blegdamsvej 17\\ DK-2100 Copenhagen\\
Denmark}
\begin{document}
\maketitle
\vfill
\noindent \textbf{Abstract.} In this paper we describe a renormalon-inspired
model of twist--4 power corrections to the light-cone pion distribution
amplitude, and compare it to the results obtained using the conformal wave
expansion. We find that the overall functional form predicted in this
renormalon model is similar to the one predicted in the standard approach to
higher twist contributions involving an expansion in conformal
operators. However, small discrepancies at the end-points allow for a
discussion on the convergence properties of the conformal expansion.
\vfill
\begin{flushleft}
NBI-HE-99-33
\end{flushleft}
\thispagestyle{empty}
\newpage
\setcounter{page}{1}
\section{Introduction}
\label{sec:intro}
Hadron distribution amplitudes play a key role in the study of hard exclusive
processes in QCD. At large $Q^2$, amplitudes for such processes can be
written as a convolution of a process dependent \emph{hard scattering
  amplitude} which is calculable in perturbation theory, and a process
independent non-perturbative \emph{distribution amplitude}, one for each
hadron participating in the process~\cite{Lepage:1980fj}. The leading
twist--2 distribution amplitude for the pion at asymptotically large $Q^2$ is
given by
\begin{eqnarray}
  \label{eq:phias}
  \phi(u)=6u(1-u),
\end{eqnarray}
to which there are corrections at finite $Q^2$ in the form of Gegenbauer
polynomials. It is, however, generally believed that there will be large
power suppressed corrections to the usual factorisation at accessible
energies. For some observables (eg.~the pion transition form factor for the
process $\pi\to\gamma^*\gamma^*$ and light-cone sum rules for the decay
$B\to\pi e \nu$), these power suppressed contributions are calculable in
terms of process independent \emph{higher twist} distribution amplitudes
coming from higher Fock states and transverse momentum of the quarks.
However, for eg.~the pion form factor one has to also take into account
so-called \emph{soft contributions} (see
eg.~\cite{Agaev:1995yf,Braun:1994ij}) that generally result in power
corrections to the process dependent hard scattering amplitudes. This paper
will not consider the problem of factorisation at the power suppressed levels
or the soft contributions. Instead, we will focus on the higher twist
distribution amplitudes from transverse momentum and Fock states with
different spin components.

% One has to distinguish between corrections from so-called
% \emph{soft contributions} coming from the end-points of momentum fraction
% integrals and \emph{higher twist corrections} coming from higher Fock states
% and transverse momentum of the quarks.  This paper will not consider the soft
% contribution. Instead, we will focus on the higher twist contributions from
% transverse momentum and Fock states with different spin components.

A gauge-invariant way to study the expansion of the pion distribution
amplitude in terms of increasing twist is to consider the vacuum-to-pion
matrix element. We parametrise it as~\cite{Braun:1990iv}
\begin{eqnarray}
  \label{eq:twistexp}
 && \bbra 0 \bar d (0) \gamma_\mu \gamma_5 [0,x]u(x)\bket {\pi^+(p)}\\
 &=& i f_\pi p_\mu \int_0 ^1  \d u e^{-iupx}\left(\phi(u) + x^2 g_1(u) +
 O(x^4)\right)\nonumber\\
& +& f_\pi \left(x_\mu - \frac{x^2 p_\mu}{px}\right) \int _0 ^1
 \d u e^{-iupx}g_2(u)+\cdots\nonumber,
\end{eqnarray}
where $\phi(u)$ is the asymptotic leading twist contribution (\ref{eq:phias})
and $[0,x]$ denotes the path-ordered gauge factor along the straight line
between $0$ and $x$
\begin{eqnarray}
  [x,y]=P\exp\left[ig\int_0^1\d t (x-y)_\mu A^\mu\left(tx+(1-t)y\right)\right].
\end{eqnarray}
$g_1(u)$ and $g_2(u)$ parametrise the twist--4 part and correspond to the
contributions from the transverse momentum of the quarks and from states of
the quarks with a different spin projection $\bar\psi\gamma_-\psi$,
respectively.

The standard procedure for estimating higher twist contributions is to
exploit a connection between $g_1(u),g_2(u)$ and the three-particle
amplitudes. This connection is derived using the QCD equations of motion and
allow one to calculate $g_1(u)$ and $g_2(u)$ through an expansion in
conformal operators of the three-particle operator.
%  which is derived using the QCD equations of motion. The three
% particle amplitudes are then expanded in terms of increasing conformal spin.
It is generally believed that the anomalous dimensions of operators increase
with increasing conformal spin, and so only a few terms in this expansion
should be needed to get reliable results. The analysis in~\cite{Braun:1990iv}
includes the contribution to twist--4 distribution amplitudes from the
operators of lowest and next-to lowest conformal spin.  In this framework
there has been recent progress in the study of pre-asymptotic behaviour of
the meson distribution amplitudes, extending the analysis to vector mesons
and including mass-corrections~\cite{Ball:1998sk,Ball:1998ff}. The results
are being extensively used, e.g.~in light-cone sum rule calculations of
B-meson decay form factors~\cite{Ball:1998kk} etc.

In the following we will describe a different approach that does not rely on
the hierarchy of anomalous dimensions and therefore allows us to get some
insight into the convergence properties of the conformal expansion. 

\section{Light cone pion distribution amplitudes}
\label{sec:lightcone}
The relation (\ref{eq:twistexp}) of the vacuum-to-pion transition amplitude
has to be understood as an operator product expansion taken at tree level.
At higher orders there will be corrections to eg.~the coefficient of
$\phi(u)$, which is taken to unity in (\ref{eq:twistexp}).  A more complete
form of (\ref{eq:twistexp}) to twist--4 accuracy is, therefore,
(gauge-factors are implicit)
\begin{eqnarray}
  \label{eq:lightconeexp}
 && \bbra 0 \bar d (0) \gamma_\mu \gamma_5 u(x)\bket {\pi^+(p)}\\
 &=& i f_\pi p_\mu \int_0 ^1  \d u e^{-iupx}
\left[\sum_{n=0} r_n(u,\mu^2) \alpha_s^{n}(\mu^2)
 \phi(u,\mu^2) 
   + x^2
  g_1(u,\mu^2) 
 \right]\nonumber\\
& +& f_\pi \left(x_\mu 
- \frac{x^2 p_\mu}{px}
\right) \int _0 ^1
 \d u e^{-iupx}g_2(u,\mu^2)\nonumber,
\end{eqnarray}
where the distribution amplitudes $\phi(u),g_1(u)$ and $g_2(u)$ are defined
as matrix elements of non\-local quark-antiquark operators at strictly
light-like separation.  The projection on the light-cone introduces
divergences, and the distribution amplitudes thus have to be renormalised. We
choose the $\overline{\mbox{MS}}$ subtraction scheme and the normalisation
scale $\mu^2=1/x^2$ so that the coefficient functions are free from
logarithms of the type $\log(x^2\mu^2)$.

It is generally believed that the perturbative series $\sum_n r_n
\alpha_s^{n}$ will have factorially divergent coefficients
$r_n$\cite{Beneke:1998ui}. This introduces ambiguities in the definition of
the sum of the series, usually referred to as \emph{infrared renormalon
  ambiguities}.  These ambiguities are artifacts of the perturbative
factorisation, and for a physically observable transition amplitude these
divergences must cancel with the corresponding \emph{ultraviolet} renormalon
ambiguities in the definition of the higher twist distribution amplitudes.
In a different language, the renormalon problem indicates a double counting
of the regions of small momenta if perturbation theory is not calculated with
an explicit IR cutoff, and the double counting is reflected by quadratic
divergences of the twist-4 operators. We refer the reader to
Ref.~\cite{Beneke:1998ui} for the thorough discussion. The higher-twist
contributions are, therefore, necessary to render the perturbative treatment
of physical observables well defined. It follows that the ambiguities of the
leading-twist perturbative series can be taken as a ``minimal model'' of
higher-twist distribution amplitudes. This we will call the \emph{renormalon
  model}.

Instead of calculating the higher-order contributions to the leading-twist
coefficient function directly, it is more instructive to consider the
ultraviolet renormalon ambiguity in twist--4 operators, which is equivalent
according to the above argument. The calculation of the ultraviolet
renormalon ambiguity in the twist--4 operators is similar to a calculation by
L.~Magnea, V.M.~Braun and M.~Beneke reported in~\cite{Beneke:1998ui}. As in
the standard approach, it is convenient to study first the three-particle
quark-antiquark-gluon operators and use the equations of motion.
Specifically, the following identities, derived from the equations of motion
for QCD string operators~\cite{Balitsky:1989bk}, can be used:
\begin{eqnarray}
  \label{eq:eomrel1}
  \frac{\partial}{\partial x_\mu}\bar d (-x) \gamma_\mu \gamma_5 u(x) &=& i
  \int _{-1}^1 \d v v\bar d(-x)x_\alpha g
  G_{\alpha\beta}(vx)\gamma_\beta\gamma_5 u(x),\\
  \partial_\mu \left\{\bar d (-x) \gamma_\mu \gamma_5 u(x) \right\} &=& i \int
  _{-1}^1 \d v \bar d(-x)x_\alpha  g  G_{\alpha\beta}(vx)\gamma_\beta\gamma_5
  u(x).
  \label{eq:eomrel2}
\end{eqnarray}
Here $\partial_\mu$ is the derivative with respect to a translation of the
nonlocal operator, while $\frac{\partial}{\partial x_\mu}$ is the derivative
with respect to the interquark separation. $G_{\alpha\beta}$ is the gluon
field strength tensor. The three-particle light-cone distribution amplitudes
on the right hand side can be parametrised as~\cite{Braun:1990iv}
\begin{eqnarray}
  \label{eq:3twist4}
&&\bbra{0}\bar d (-x) \gamma_\mu\gamma_5
G_{\alpha\beta}(vx)u(x)\bket{\pi^+(p)}\nonumber\\
&=&p_u(p_\alpha x_\beta-p_\beta x_\alpha)\frac 1 {p x}f_\pi\int \D \alpha_i
\Phipar (\alpha_i) e^{-ipx(\alpha_1-\alpha_2+v\alpha_3)}\\
&+&\left[
p_\beta\left(g_{\alpha\mu}-\frac{x_\alpha p_\mu}{px}\right)
-p_\alpha\left(g_{\beta\mu}-\frac{x_\beta p_\mu}{px}\right)
\right]f_\pi\int \D \alpha_i
\Phiperp(\alpha_i)e^{-ipx(\alpha_1-\alpha_2+v\alpha_3)}, \nonumber
\end{eqnarray}
The operator on the left hand side contains contributions of both twist--3
and twist--4, but the matrix element of the twist--3 part vanishes. The pion
decay constant $f_\pi$ is defined through
\begin{eqnarray}
  \label{eq:piondecayconstant}
  \bra 0 \bar u (0) \gamma_\mu\gamma_5 d(0)\ket {\pi^+(P)} = i f_\pi P_\mu,
\end{eqnarray}
and the integration measure on the right hand side is defined as
\begin{eqnarray}
  \label{eq:intmeasure}
  \int \D \alpha_i = \int _0 ^1 \d \alpha_1\d\alpha_2\d\alpha_3 \delta
  (1-\alpha_1-\alpha_2-\alpha_3).
\end{eqnarray}
By sandwiching (\ref{eq:eomrel1}) and (\ref{eq:eomrel2}) between the vacuum
and pion state it is now possible to derive the following relations between
the parameterisations of the twist--4 distribution amplitudes and the three
particle operator~\cite{Braun:1990iv}
\begin{eqnarray}
  \label{eq:grel}
  g_2(u)&=&\int_0^u\d\alpha_1\int_0^{\bar u}\d \alpha_2 \frac 1 {\alpha_3}
  [2\Phiperp - \Phipar](\alpha_1,\alpha_2,\alpha_3),\\
  g_1(u)+\int_0^u\d v g_2(v)&=&\frac 1 2 \int_0^u\d\alpha_1\int_0^{\bar u
  }\d\alpha_2 \frac 1 {\alpha_3^2}(\bar u \alpha_1 - u \alpha_2)[2\Phiperp -
  \Phipar](\alpha_1,\alpha_2,\alpha_3). \nonumber
\end{eqnarray}
Here $\alpha_3=1-\alpha_1-\alpha_2$ and $\bar u =1-u$. In the following
analysis we will calculate an explicit expression for the special combination
$[2\Phiperp-\Phipar](\alpha_1,\alpha_2,\alpha_3)$ in the renormalon model
and restore $g_1(u)$ and $g_2(u)$ from these expressions.

\section{The Renormalon Calculation}
\label{sec:renormalon}
It is generally believed that although the perturbative coefficients for QCD
perturbative series show signs of factorial divergence, the series are
\emph{asymptotic series} for which we can assign a sum using the Borel
transformation~\cite{Beneke:1998ui}. The Borel transform of a perturbative
series
\begin{eqnarray}
  \label{eq:series}
  R(a)=r_0a+r_1a^2+r_2a^3+\cdots = \sum_{n=0}^\infty r_n a^{n+1}
\end{eqnarray}
is defined as
\begin{eqnarray}
  \label{eq:boreltransform}
  B[R](t)=\sum_{n=0}^\infty t^n \frac {r^n}{n!}
\end{eqnarray}
If $B[R](t)$ is well-defined on the positive real axis and is integrable we
may define the \emph{Borel sum} of the series as
\begin{eqnarray}
  \label{eq:borelsum}
  \tilde R = \int_0^\infty\d t e^{-t/a} B[R](t),
\end{eqnarray}
with a series expansion identical to that of $R$. For convergent series, the
Borel sum is therefore identical to the sum of the original series, and we
can regard the Borel sum as an extension of the definition of the sum of a
series to divergent, asymptotic series.

However, for a QCD observable $R$ the Borel transform $B[R]$ has
singularities on the real axis, leading to ambiguities in the definition of
the integral (\ref{eq:borelsum}). These so-called \emph{renormalons} are a
direct result of the divergence of the original series, and as already
mentioned the residues of the renormalon poles are taken as an estimate of
the non-perturbative contribution to the observable~\cite{Beneke:1998ui}. For
QCD observables, one finds that there are infrared renormalons stemming from
the low-momentum region of higher order loop contributions, situated at
$w=t/\beta_0=1,2,\ldots$. Here $\beta_0$ is the first $\beta$--function
coefficient $\frac{\d\alpha_s}{\d\ln\mu^2}=-\beta_0\alpha_s^2+\cdots$.

As seen from (\ref{eq:borelsum}), the contribution from a pole situated at
$w_i$ in the Borel plane to the observable will be suppressed by
$\exp(-w_i\beta_0/\alpha_s)$. As the QCD coupling (at one-loop approximation)
fulfils $1/\alpha_s=\beta_0\ln(Q^2/\Lambda^2)$, we will need the IR
renormalon at $w=1$ to obtain the $1/Q^2$ twist--4 corrections.

It is certainly impossible to calculate the residue of the Borel transform at
a given pole exactly, since that would require full knowledge of all the
perturbative coefficients in the series. Instead we will apply the so-called
\emph{Naive non-Abelianization}
(NNA)~\cite{Broadhurst:1995se,Beneke:1995qe,Ball:1995ni,Lovett-Turner:1995ti}
in which the coefficients are approximated by their leading $\beta_0$ part.
The NNA starts from the approximation of the perturbative coefficients with
the part coming from diagrams with one string of fermion bubbles inserted to
a gluon line. In this way the leading $N_f$ piece ($N_f$ being the number of
quark flavours) is found. The full one-loop $\beta$ function behaviour is
then restored by substituting for $N_f$ the scaled $\beta$-function
coefficient $N_f-\frac {33} 2$. This results in a gauge-invariant
approximation of the diagrams to all orders in the
coupling~\cite{Beneke:1998ui}. The positions of the poles in the Borel plane
do not depend on the higher $\beta$-function coefficients and so are
unchanged in the NNA compared to the full theory. On the other hand, the
residues change and one actually finds that also the nature of the
divergences may change. In this way, the pole-divergence of the Borel
transform at $w=1$ in the NNA is actually the end-point of a branch cut in
the full theory.

The NNA is useful for doing calculations, since in this approximation the
Borel transformation applies to the modified gluon propagator itself rather
than the diagram as a whole. The Borel-transformed effective gluon propagator
in the Lorentz gauge is given by~\cite{Beneke:1998ui}
\begin{eqnarray}
  \label{eq:borelgluonprop}
  B[\alpha_s(\mu^2)D_{\mu\nu}](w)&=&\frac{-i}{k^2}\left(g_{\mu\nu}-\frac{k_\mu
  k_\nu}{k^2} \right) \left(-\frac{\mu^2}{k^2}e^{-C}\right)^w,
\end{eqnarray}
where the $k_\mu k_\nu$-terms can be neglected for gauge-invariant
quantities. $\mu$ is the renormalisation scale and $C$ is a renormalisation
scheme-dependent constant. In the $\overline{\mbox{MS}}$-scheme $C=-5/3$. It
is noted that the only essential difference between (\ref{eq:borelgluonprop})
and the original gluon propagator is the power of $k^2$ in the denominator,
and therefore a calculation of the Borel transform of a diagram in the NNA
to all orders in the coupling is no more difficult than the calculation of
the same diagram with a simple gluon propagator.

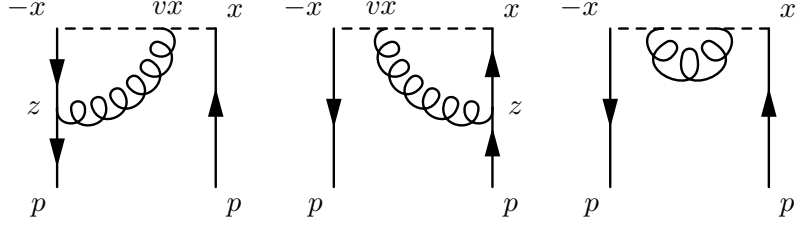
\begin{figure}[tbp]
  \begin{center}
    \begin{fmffile}{graph1}
  \fmfframe(15,0)(15,0){
  \begin{fmfgraph*}(60,60)
    \fmfstraight
    \fmfleft{l1,l2,l3}
    \fmfright{r1,r2,r3}
    \fmf{phantom}{r1,r2}
    \fmf{phantom}{r2,r3}
    \fmf{phantom}{l1,l2}
    \fmf{phantom}{l2,l3}
    \fmffreeze
    \fmf{dashes}{r3,v1,v2,l3}
    \fmffreeze
    \fmf{fermion}{l3,l2,l1}
    \fmf{fermion}{r1,r3}
    \fmf{gluon,right=0.4}{l2,v1}
    \fmflabel{$p$}{l1}
    \fmflabel{$p$}{r1}
    \fmflabel{$z$}{l2}
    \fmflabel{$-x$}{l3}
    \fmflabel{$x$}{r3}
    \fmflabel{$vx$}{v1}
%    \fmfv{decor.shape=circle,decor.filled=empty}{v1}
  \end{fmfgraph*}}
\end{fmffile}
\begin{fmffile}{graph2}
  \fmfframe(15,0)(15,0){
  \begin{fmfgraph*}(60,60)
    \fmfstraight
    \fmfleft{l1,l2,l3}
    \fmfright{r1,r2,r3}
    \fmf{phantom}{r1,r2}
    \fmf{phantom}{r2,r3}
    \fmf{phantom}{l1,l2}
    \fmf{phantom}{l2,l3}
    \fmffreeze
    \fmf{dashes}{r3,v1,v2,l3}
    \fmffreeze
    \fmf{fermion}{l3,l1}
    \fmf{fermion}{r1,r2,r3}
    \fmf{gluon,right=0.4}{v2,r2}
    \fmflabel{$p$}{l1}
    \fmflabel{$p$}{r1}
    \fmflabel{$z$}{r2}
    \fmflabel{$-x$}{l3}
    \fmflabel{$x$}{r3}
    \fmflabel{$vx$}{v2}
  \end{fmfgraph*}}
\end{fmffile}
\begin{fmffile}{graph3}
  \fmfframe(15,0)(15,0){
  \begin{fmfgraph*}(60,60)
    \fmfstraight
    \fmfleft{l1,l2,l3}
    \fmfright{r1,r2,r3}
    \fmf{phantom}{r1,r2}
    \fmf{phantom}{r2,r3}
    \fmf{phantom}{l1,l2}
    \fmf{phantom}{l2,l3}
    \fmffreeze
    \fmf{dashes}{r3,v1,v2,l3}
    \fmffreeze
    \fmf{fermion}{l3,l1}
    \fmf{fermion}{r1,r3}
    \fmf{gluon,right}{v2,v1}
    \fmflabel{$p$}{l1}
    \fmflabel{$p$}{r1}
    \fmflabel{$-x$}{l3}
    \fmflabel{$x$}{r3}
  \end{fmfgraph*}}
\end{fmffile}
    \caption{The twist--4 contributions to the renormalisation of the
      three-particle operator~(\ref{eq:3twist4}). The third diagram is
      scale-less and vanishing~\cite{Beneke:1998ui}.}
    \label{fig:feynman}
  \end{center}
\end{figure}

We start by noting that to project out the the sought combination $[2\Phiperp
- \Phipar]$, we have to contract the three-particle operator in
(\ref{eq:3twist4}) with $(-x^\alpha g^{\mu\beta})$. We want to calculate the
leading ultraviolet renormalon singularity in the matrix element of this
operator at $w=1$. Since this singularity comes from contributions of large
momenta, the particular external states do not matter and it is sufficient to
evaluate the contribution of the diagrams in Fig.~\ref{fig:feynman} between
quark states.

In the Borel representation the matrix element of the projected operator in
(\ref{eq:3twist4}) between quark states of momentum $p$ is given by
\begin{eqnarray}
  \label{eq:melementfreequark}
 &&  \bbra p \bar d (-x) \gamma_\mu\gamma_5 G_{\alpha\beta}(vx)u(x) x^\alpha
  g^{\mu\beta} \bket p\nonumber\\
  &=& i 4\pi \cf \left( \mu ^ 2 e^{-C} \right)^w \bar d (p) e^{-2ipx}\left
  (\gud\lambda\mu x^\nu - x_\mu g^{\lambda\nu}\right)\nonumber\\
&\cdot& \left\{ \gamma^\mu\gamma^\sigma\gamma_\lambda\gamma_5 \phsiint k
  e^{-i\tilde v k x} \frac{k_\nu
  (p+k)_\sigma}{(-k^2)^{w+1}\left[-(p+k)^2\right]} \right.\nonumber\\
&&+
\left. \gamma_\lambda\gamma_5\gamma^\sigma\gamma^\mu\phsiint k e^{i\bar v kx}
  \frac{k_\nu(p-k)_\sigma} {(-k^2)^{w+1}\left[-(p-k)^2\right]}
\right \} u(p),
\end{eqnarray}
where $\bar v=1-v$.  For $x$ strictly light-like, we find for the
Borel-transformed operators in a neighbourhood of $w=1$
\begin{eqnarray}
  \label{eq:twist42relation}
&&  \bar d (-x) \gamma_\mu\gamma_5 G_{\alpha\beta}(vx)u(x) x^\alpha
  g^{\mu\beta} \nonumber\\
&=\atop {\mbox{\tiny quad. diverg.}}& 2i(4\pi) \cf \mu^2 e^{-C} \frac 1 {32\pi^2} \frac 1 {w-1} \nonumber \\
&&\cdot \int _0 ^1 \d
  \alpha \alpha \left\{ \bar d \left[-x\right]\gamma_5 \ds x\ u\! \left
  [x(1-\alpha\bar v)\right] - \bar d \left[-x(1-\alpha\tilde v) \right]
  \gamma_5 \ds x\ u\! \left [x\right] \right\}.
\end{eqnarray}
As already mentioned, the quadratic ultraviolet divergence of the residue of
the pole is independent of the external states, which allows us to write the
relation as an operator relation. Note by comparing
(\ref{eq:twist42relation}) and (\ref{eq:twistexp}) that we have written the
twist--4 operator as an integral over leading twist contributions. By taking
the residue at $w=1$ of the pion to vacuum transition amplitude of the
twist--4 operator we find
\begin{eqnarray}
\label{eq:phif}
[2\Phiperp - \Phipar](\alpha_i) &=&
-\cf \mu^2 e^{-C} \frac 1 {4\pi} \left\{\frac{\alpha_3}{(1-\alpha_2)^2}\phi(1-\alpha_2) - \frac {\alpha_3}
  {(1-\alpha_1) ^2}\phi(\alpha_1)\right\},
\end{eqnarray}
Choosing the asymptotic expression for the pion distribution amplitude
$\phi(u)$ in (\ref{eq:phif}), and inserting this result in (\ref{eq:grel}) we
obtain the following explicit expressions for the distribution amplitudes
$g_1(u),g_2(u)$:
\begin{eqnarray}
  \label{eq:g12functions}
  g_2(u)&=& N (u \log u-\bar u \log \bar u)\\
  g_1(u)&=& N \frac 1 4 \bigg( - 2 \bar u u +2 \log u \log \bar u + 2\bar u \log
  \bar u +2 u \log u \nonumber\\
&& -\bar u \log ^2 \bar u -u \log ^2 u - 2u\li\left(\frac
  {-\bar u }u\right) -2\bar u \li \left(\frac{-u}{\bar u}\right)\bigg).\\
N&=&\frac 3 {2\pi}\cf\mu^2e^{-C}\nonumber
\end{eqnarray}
with $\li(z)$ the \emph{di-logarithm}.

\section{Discussion}
\label{sec:discussion}
Unfortunately, the overall normalisation of the distribution amplitudes
$g_1(u),g_2(u)$ is not accessible from the renormalon model and has to be
found from eg.~the QCD sum rule method. This is also the case for renormalon
models of DIS structure functions and the conformal approach to pion
distribution amplitudes.

However, the normalisation factor N found using the renormalon residue can be
taken to indicate the order of magnitude of the twist-4 correction. Using
$\mu=\Lambda=0.250$GeV we get $N\approx 0.091$GeV$^2$. Thereby we find
$\int_0^1 g_1(u)\d u=0.0088$GeV$^2$ which has to be compared with the result
obtained by normalising the conformal results using light-cone QCD sum rules.
Using the functional form and normalisation for the conformal approach of
Ref.~\cite{Braun:1990iv} we find $\int_0^1 g_1(u)\d u=0.028$GeV$^2$. For the
first moment of the distribution amplitude $g_2$ we find $\int_0^1\d u(2u-1)
g_2(u)=0.0051$GeV$^2$ and $\int_0^1\d u(2u-1) g_2(u)=0.011$GeV$^2$ for the
renormalon and conformal approach respectively.

\begin{figure}[htbp]
  \begin{center}
    \epsfig{file=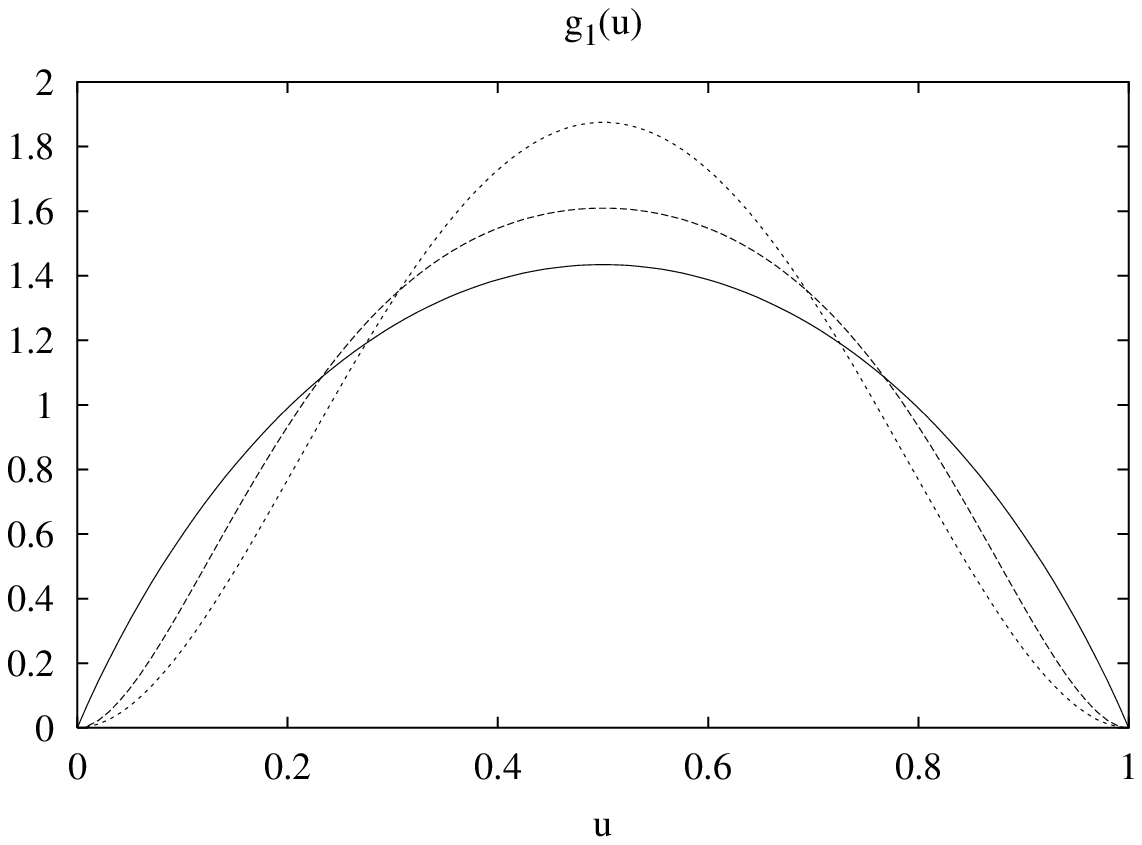,width=7.5cm} \hspace{0.2cm}
    \epsfig{file=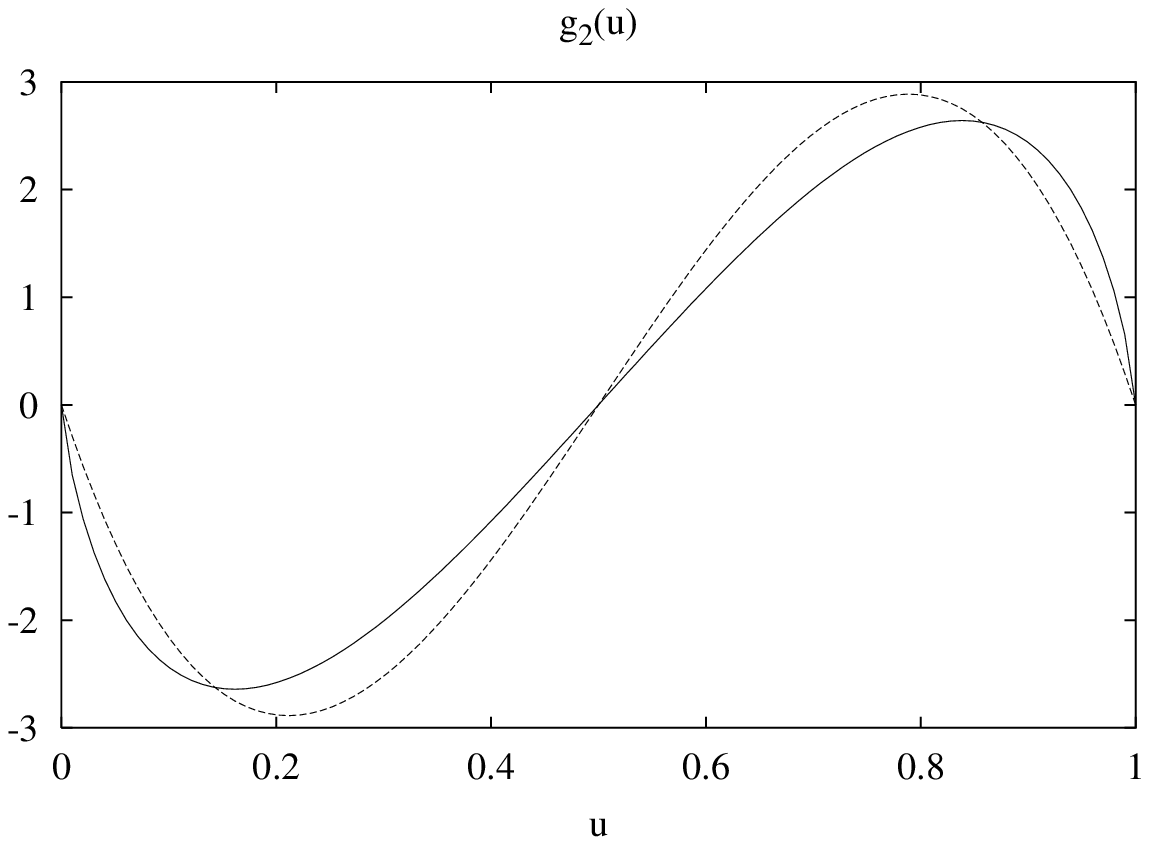,width=7.5cm}
    \caption{$g_1(u)$ (left) and $g_2(u)$ (right) calculated in both the
      renormalon model (line) and the conformal wave expansion of
      \cite{Braun:1990iv} (dots for the lowest conformal spin result and
      dashes for the result including next-to-lowest conformal spin
      operators). The distribution amplitudes have been normalised so that
      $\int_0^1\d u g_1(u)=1$ and $\int_0^1\d u(2u-1) g_2(u)=1$.}
    \label{fig:gs}
  \end{center}
\end{figure}

It is more interesting that the shapes of the predicted wave amplitudes are
close to the ones obtained in the conformal approach of~\cite{Braun:1990iv},
see Fig.~\ref{fig:gs}. The functions have been normalised so that $\int_0^1\d
u g_1(u)=1$ and $\int_0^1\d u(2u-1) g_2(u)=1$. The similarity is encouraging,
since it indicates that the approximations are well understood.  The somewhat
broader shape of the distribution amplitude $g_1(u)$ in the renormalon model
compared to the conformal expansion is explained by the contributions of
higher conformal spin operators.  The anomalous dimensions of operators are
ignored in the NNA approximation, so that it effectively corresponds to the
summation of the conformal series to all orders, including, however, all
kinematic suppression factors. The renormalon model, therefore, pretends to
make a simple estimate of higher orders in the conformal expansion, and the
main result of our study is that such contributions seem to be comfortably
small, apart from in the end-point regions.  The asymptotic behaviour of the
renormalon model prediction for the distribution amplitude $g_2(u)$ at
$x\to0,1$ coincides with that of the lowest conformal spin contribution to
$g_2(u)$.  At the same time, the behaviour of $g_1(u)$ is different. The
renormalon result behaves as $u$ for $u\to0$ compared to the $u^2$ behaviour
of each conformal operator, indicating that the series in conformal operators
is not converging uniformly. This feature will have an effect on observables
sensitive to these regions like $B$-meson decay form
factors~\cite{Ball:1998kk}.

\section{Acknowledgements}
\label{sec:acknowledgements}
The author is grateful to V.M.~Braun for introducing him to the subject and
for countless valuable discussions. Also, the Lørup Foundation is
acknowledged for financial support for a stay in Regensburg during the final
stages of this work and for participation in the workshop ``Light-Cone
Wave Functions in QCD'' in Regensburg.

\bibliographystyle{unsrt}
\bibliography{database}

\begin{thebibliography}{10}

\bibitem{Lepage:1980fj}
G.~Peter Lepage and Stanley~J. Brodsky.
\newblock Exclusive processes in perturbative quantum chromodynamics.
\newblock {\em Phys. Rev.}, D22:2157, 1980.

\bibitem{Agaev:1995yf}
S.~S. Agaev.
\newblock Electromagnetic pion form-factor in the context of the principal
  value method.
\newblock {\em Phys. Lett.}, B360:117--122, 1995.

\bibitem{Braun:1994ij}
Vladimir Braun and Igor Halperin.
\newblock Soft contribution to the pion form-factor from light cone {QCD} sum
  rules.
\newblock {\em Phys. Lett.}, B328:457--465, 1994.

\bibitem{Braun:1990iv}
V.~M. Braun and I.~E. Filyanov.
\newblock Conformal invariance and pion wave functions of nonleading twist.
\newblock {\em Z. Phys.}, C48:239--248, 1990.

\bibitem{Ball:1998sk}
Patricia Ball, V.~M. Braun, Y.~Koike, and K.~Tanaka.
\newblock Higher twist distribution amplitudes of vector mesons in {QCD}:
  Formalism and twist three distributions.
\newblock {\em Nucl. Phys.}, B529:323, 1998.

\bibitem{Ball:1998ff}
Patricia Ball and V.~M. Braun.
\newblock Higher twist distribution amplitudes of vector mesons in {QCD}: Twist
  - 4 distributions and meson mass corrections.
\newblock {\em Nucl. Phys.}, B543:201, 1999.

\bibitem{Ball:1998kk}
Patricia Ball and V.~M. Braun.
\newblock Exclusive semileptonic and rare {B} meson decays in {QCD}.
\newblock {\em Phys. Rev.}, D58:094016, 1998.

\bibitem{Beneke:1998ui}
M.~Beneke.
\newblock Renormalons.
\newblock {\em Phys. Rept.}, 317:1, 1999.

\bibitem{Balitsky:1989bk}
I.~I. Balitsky and V.~M. Braun.
\newblock Evolution equations for {QCD} string operators.
\newblock {\em Nucl. Phys.}, B311:541--584, 1989.

\bibitem{Broadhurst:1995se}
D.~J. Broadhurst and A.~G. Grozin.
\newblock Matching {QCD} and {HQET} heavy - light currents at two loops and
  beyond.
\newblock {\em Phys. Rev.}, D52:4082--4098, 1995.

\bibitem{Beneke:1995qe}
M.~Beneke and V.~M. Braun.
\newblock Naive non{A}belianization and resummation of fermion bubble chains.
\newblock {\em Phys. Lett.}, B348:513--520, 1995.

\bibitem{Ball:1995ni}
Patricia Ball, M.~Beneke, and V.~M. Braun.
\newblock Resummation of $(\beta_0 \alpha_s)^n$ corrections in {QCD}:
  Techniques and applications to the $\tau$ hadronic width and the heavy quark
  pole mass.
\newblock {\em Nucl. Phys.}, B452:563--625, 1995.

\bibitem{Lovett-Turner:1995ti}
C.~N. Lovett-Turner and C.~J. Maxwell.
\newblock All orders renormalon resummations for some {QCD} observables.
\newblock {\em Nucl. Phys.}, B452:188--212, 1995.

\end{thebibliography}

\end{document}